\documentclass[journal,final]{IEEEtran}

\ifCLASSINFOpdf
  \usepackage[pdftex]{graphicx}
\usepackage{epsfig}    
\usepackage{caption}
\usepackage{subcaption}
\usepackage[dvipsnames]{xcolor}
\usepackage{amssymb}
\usepackage{algorithmic,algorithm}
\usepackage{color}
\usepackage{multicol}
\usepackage[multiple]{footmisc}
\usepackage[affil-it]{authblk} 
\usepackage{booktabs}
\usepackage{siunitx}
\usepackage{tabularx}
\usepackage{etoolbox}
\usepackage{arydshln}
\usepackage{rotating}
\usepackage{afterpage}
\usepackage{array}
\usepackage{subcaption}
\usepackage{cite}
\usepackage{color}
\usepackage{amsmath,amssymb}
\usepackage{mathtools}
\newtheorem{assumption}{\textbf{Assumption}}

\newtheorem{proposition}{\textbf{Proposition}}
\usepackage{color,soul}
\ifx\pdfoutput\undefined
\else \usepackage[pdftex]{graphicx}
\usepackage{epstopdf}
\makeatletter
\let\MYcaption\@makecaption
\makeatother
\usepackage[font=footnotesize]{subcaption}
\usepackage[font=footnotesize]{caption}
\makeatletter
\let\@makecaption\MYcaption
\makeatother

\fi

\makeatletter
\patchcmd{\@maketitle}{\LARGE \@title}{\fontsize{24}{19.2}\selectfont\@title}{}{}
\makeatother

\begin{document}
\title{Coordinated Control of Energy Storage in Networked Microgrids under Unpredicted Load Demands}

\author{Md Tanvir Arafat~Khan,~\IEEEmembership{Student Member,~IEEE,}
        Rafael~Cisneros,
        Aranya~Chakrabortty,~\IEEEmembership{Senior Member,~IEEE,}
        Iqbal~Husain,~\IEEEmembership{Fellow,~IEEE}
\thanks{M. T. A. Khan, R. Cisneros, A. Chakrabortty and I. Husain are with North Carolina State University, Raleigh, NC 27606, USA. E-mails:  mtkhan[rcisner,achakra2,ihusain2]@ncsu.edu.}
\thanks{This work is supported by the National Science Foundation, under Award No. EEC-0812121 for the FREEDM Engineering Research Center.}
\vspace{-2em} }
\makeatletter
	\renewcommand*\env@matrix[1][c]{\hskip -\arraycolsep
	\let\@ifnextchar\new@ifnextchar
 	\array{*\c@MaxMatrixCols #1}}
\makeatother

\maketitle

\begin{abstract}
\boldmath
In this paper a nonlinear control design for power balancing in networked microgrids using energy storage devices is presented. Each microgrid is considered to be interfaced to the distribution feeder though a solid-state transformer (SST). The internal duty cycle based controllers of each SST ensures stable regulation of power commands during normal operation. But problem arises when a sudden change in load or generation occurs in any microgrid in a completely unpredicted way in between the time instants at which the SSTs receive their power setpoints. In such a case, the energy storage units in that microgrid must produce or absorb the deficit power. The challenge lies in designing a suitable regulator for this purpose owing to the nonlinearity of the battery model and its coupling with the nonlinear SST dynamics. We design an input-output linearization based controller, and show that it guarantees closed-loop stability via a cascade connection with the SST model. The design is also extended to the case when multiple SSTs must coordinate their individual storage controllers to assist a given SST whose storage capacity is insufficient to serve the unpredicted load. The design is verified using the IEEE 34-bus distribution system with nine SST-driven microgrids.
\end{abstract}
\section{Introduction}
\begin{figure}[t!]
	\includegraphics[width=0.97\columnwidth]{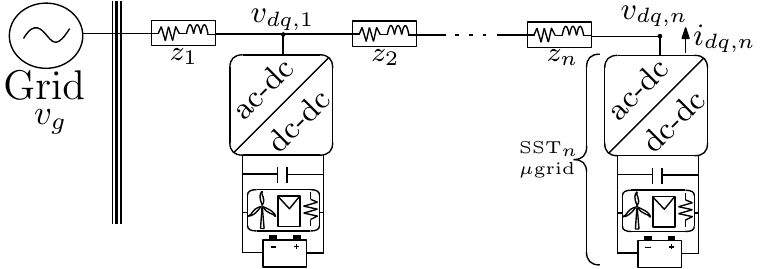}
	\caption{Radial topology: each SST-driven microgrid is composed of renewable resources, an energy storage unit and a load.}\label{top}
\end{figure}

\IEEEPARstart{I}n recent years power engineers have started visiting the concept of networked microgrids \cite{intro,intro2}, where individual microgrids are coordinated to create convenient electrical topologies that guarantee reliable flow of power from one part of the grid to another, especially during emergency scenarios. An excellent resource for sensing and controlling such power flows is a solid-state transformers (SST) \cite{huang2}. A schematic diagram of a radially networked microgrid, where each individual microgrid is interfaced with the distribution feeder through a SST, is shown in Fig \ref{top}. The SST consists of three power electronic converter stages---namely, rectifier, dual-active bridge, and inverter, which in turn are connected to AC and DC generators (for example, wind and solar PV), AC and DC loads, and most importantly a DC energy storage. The circuit diagram of a SST with these three stages is shown in Fig. \ref{fig1}.

Power balancing mechanisms for these types of networked systems typically consist of two steps. First, a supervisory controller, commonly referred to as an intelligent energy management (IEM) at the distribution substation, predicts the load for each microgrid fifteen to twenty minutes ahead of time, solves power flow, and generates the voltage and current setpoints for each SST. When the loads change at the scheduled instants of time, the internal duty cycle based controller in the rectifier circuit of the SST (referred to as an intelligent power management (IPM) controller) gets triggered, and drives the steady-state voltages and currents to the respective setpoints using available power generation from wind and solar PV. A challenge, however, arises when any load changes significantly in between the scheduled instants of IEM commands in an unforeseen and unpredicted way. In such a case, the battery of the SST must instantaneously trigger to produce or absorb the deficit power. Appropriate control systems with fast tracking properties  need to be designed for this purpose. 
While several papers in recent literature have reported such battery controls \cite{intro3,intro4,intro5}, most of them are based on simplified linear (or linealized) models of batteries that lack analytical guarantees of stability margins that can be achieved in realistic nonlinear models. Moreover, when the energy storage system is  connected to the rest of the microgrid, the nonlinear dynamics coupling both may be a source of instability as well. Therefore, stable operation of the entire microgrid needs to be established. The problem becomes even more complicated when multiple microgrids need to coordinate the control actions to serve a given SST whose storage capacity is insufficient to serve its unpredicted load.

In this paper, a nonlinear control design is proposed for solving this tracking control problem. Each energy storage unit is operated in controlled-current mode with its reference current set such that the deficit power between generation and load is driven to zero autonomously within each microgrid, thereby maintaining power balance in the network.When the deficit cannot be autonomously supported,  balance is maintained collectively via co-ordination between the battery controllers. Since it allows for regulation of power flows in each microgrid, the proposed control scheme may be viewed either as \textit{tertiary} control following the terminology used in \cite{intro}, or \textit{secondary} control following that in \cite{seco}. Furthermore,
the controller is based on the input-output linearization method \cite{isidori}. This control technique allows us to achieve an exponentially
stable tracking error. It is shown that the system, however, is not completely input-output feedback linearizable as a result of which the stability of the residual dynamics needs to be established. It is proved that the currents and the voltages of the SST remain bounded while the tracking error goes to zero. Also, the margins are provided for the stable operation of each storage unit. Finally, the stability proof is  extended to the multiple SST case. Power sharing algorithms are also provided for situations where the slack created from the load in one microgrid can be cooperatively fulfilled.
\begin{table}[b]
	\caption{Nomenclature}\label{table1}
	\centering
	\begin{center}
		\footnotesize
		\resizebox{\linewidth}{!}{
			\begin{tabular} {p{0.7cm}p{2cm}p{0.35cm}p{2cm}p{0.3cm}p{2.05cm}}
				\multicolumn{6}{c} {System Variables}\\
				\midrule 
				\multicolumn{2}{c}{Rectifier}&\multicolumn{2}{c}{DAB Converter}&\multicolumn{2}{c}{Storage}\\
				\midrule
				$i_{d},i_{q}$ &$d,q$-axis current & $v_h$& High voltage bus& $v_{in}$& Conv. input volt. \\
				$v_{f}$ & Filter cap. volt. & $v_l$ & Low voltage bus & $v_o$& Conv. outp. volt. \\
				$\xi_{1},\xi_{2}$ & Voltage control  &  $\xi_4$ &  Low voltage ctrl.  &$v_b$ & Battery voltage\\
				$\xi_{3}$ & $q$-axis control  & $I_{dab}$ & Net $\mu$grid current &$\phi$ &Conv. phase shift\\
				$d_1$&   $d$-axis crtl. input &  $\phi_s$& Phase shift ratio & $I_b$& Output current\\
				$d_2$& $q$-axis ctrl. input& & & &\\\\
				\multicolumn{6}{c}{System Parameters}\\
				\midrule
				\multicolumn{2}{c}{Rectifier}& \multicolumn{2}{c}{DAB Converter}& \multicolumn{2}{c}{Storage}\\
				\midrule  
				$C_f$ & Filter capacitor & $C_h$& High volt. cap. & $C_{in}$& Conv. input cap.  \\
				$L_f$ & Filter inductor & $L_s$ & Transf. inductor & $C_{o}$&Conv. output cap.\\
				$r_f$ & Filter resistor &   $C_l$& Low volt. cap.& $L_b$ & Conv. transf. ind. \\
				$k_1$-$k_6$ & Controllers gains & $r_h$& Input resistor & $r_{in}$& Input resistance \\
				$\omega$&Line frequency  &$n_s$& Transf. ratio &$r_{o}$& Output resistance \\ &  &  $f_s$&Switch. freq.&$n_b$& Conv. transf. ratio \\&& $k_7,k_8$& Controllers gains& $f_b$& Switch. freq. \\\\
				\multicolumn{6}{c}{System References}\\
				\midrule
				\multicolumn{2}{c}{Rectifier}& \multicolumn{2}{c}{DAB Converter}&\multicolumn{2}{c}{Storage}\\
				\midrule
				$i_{d}^\star,i_q^\star$& $d,q$-axis current  & $v_{l}^\star$& Low DC voltage & $I_b^r$& Output current\\
				$i_{f}^\star$& DC rect. voltage  & &\\
		\end{tabular}}
	\end{center}
\end{table}
 The rest of the paper is structured as follows. In Section \ref{model}, the model of a SST driven microgrid is provided. In Section \ref{control},  the proposed controller is derived and the stable operation of the closed-loop is proven for the multi-SST system. In Sections \ref{powerd} and \ref{sim}, power sharing methods are introduced. Section VI assesses the proposed controller using a nine-SST distribution system. Conclusions are drawn in Section \ref{conclusi}.
\section{Microgrid Model of the SST-driven System}\label{model}

\begin{figure}[t]
	\centering
	\includegraphics[width=1\columnwidth]{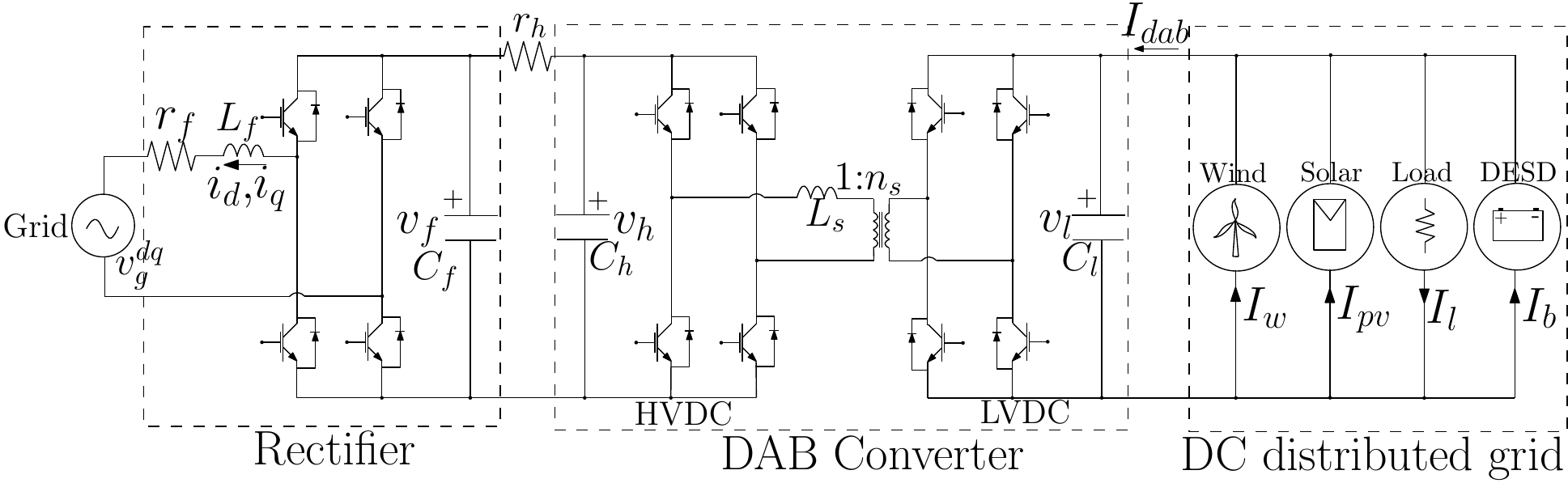}
	\caption{SST based distribution system.\label{fig1}}
\end{figure}

The microgrid model for our study is considered to be interfaced with the distribution feeder through a SST, as shown in Fig. \ref{top}. We consider operation of the microgrid in grid-connected mode, whereby the $d$ and $q$ axes voltages of the grid $v_d$ and $v_q$ act as excitation sources for the SST circuit, as shown in Fig. \ref{fig1}. Each SST consists of a front-end rectifier stage, which converts high voltage AC to high voltage DC, a dual-active bridge (DAB) stage, which converts high DC voltage to a low DC voltage to be used for DC distribution segment, and a voltage-source inverter, which converts low DC voltage to low single-phase AC voltage to be used for AC distribution segments \cite{kara}. The rectifier is responsible for maintaining the high voltage (HV) DC---i.e., the input voltage of the DAB stage. The DAB converter regulates the low voltage (LV) DC bus. Wind power generators, solar/photovoltaic generators,  loads, and an energy storage unit are assumed to be connected at the back-end of the SST. Since storage will always be connected to the DC bus, the inverter stage of the SST is not considered in the present study. Also,  the network consist of $n$  microgrids connected over a radial topology, as shown in Fig. \ref{top}---see also \cite{khan}. In the following, we briefly recall the state equations for each SST stage, keeping details to only as much as we would need to design our controller in Section \ref{control}.
\subsection{Rectifier} 

The $dq$-dynamics of the rectifier in the $i^{th}$ SST (for $i=1,\cdots,n$) are\footnote{To avoid stacking, the argument $t$ in time-varying variables is omitted when clear from the context.}%
\begin{align}
\dot{i}_{di}=&-\frac{r_{fi}}{L_{fi}}i_{di}+\omega_{i}i_{qi}+\frac{1}{L_{fi}}\,d_{1i}\,v_{fi}  -\frac{1}{L_{fi}}v_{di} \label{x1}
\end{align}
\begin{align}
\dot{i}_{qi}=&-\omega_{i}i_{di}-\frac{r_{fi}}{L_{fi}}i_{qi}+\frac{1}{L_{fi}}d_{2i}v_{f} -\frac{1}{L_{fi}}v_{qi}  \label{x2}\\
\dot v_{fi}=&-\frac{1}{2C_{fi}}(d_{1i}i_{di}+d_{2i}i_{qi})-\frac{v_{fi}-v_{hi}}{C_{fi}}+\frac{1}{C_{fi}}\left(d_{1i}i_{di}\right.\nonumber\\&\hspace{0.1cm}+d_{2i}i_{qi})\cos 2\theta_i+\frac{1}{C_{fi}}(d_{1i}i_{qi}+d_{2i}i_{di})\sin 2\theta_i\label{vf}\\
\dot{\xi}_{1i}=&v_{fi}^\star-v_{fi}\label{xi1}\\
\dot{\xi}_{2i}=&k_{1i}(v_{fi}^\star-v_{fi})+k_{2i}\xi_{1i}-i_{di}\label{xi2}\\
\dot{\xi}_{3i}=&i_{qi}^\star-i_{qi}\label{xi3} 
\end{align}
where 
\begin{align}
v_{di}=&v_g^d+r_i\sum_{j=1}^{n}i_{dj}-x_i\sum_{j=1}^{n}i_{qj}\\
v_{qi}=&v_g^q+r_i\sum_{j=1}^{n}i_{dj}+x_i\sum_{j=1}^{n}i_{qj}
\end{align}
and
\begin{align}
d_{1i}=&k_{4i}[k_{1i}(v_{fi}^\star-v_{fi})+k_{2i}\xi_{1i}-i_{di}]+k_{3i}\xi_{2i},  \label{d1}\\
d_{2i}=&k_{5i}(i_{qi}^\star-i_{qi})+k_{6i}\xi_{3i}. \label{d2}
\end{align}

\noindent The physical meanings of all the states and parameters are listed in Table \ref{table1}. Here, $v_{g}^d$ and $v_{g}^q$ are the $d$-axis and $q$-axis input grid voltage, $z_{i}=r_i+jx_i$ is the line impedance linking the $i^{th}$ SST to its aforegoing neighbor (see  Fig. \ref{top}).
\subsection{Dual Active Bridge Converter}
The second stage of a SST is the dual-active bridge (DAB) DC-DC converter. A controller is used to regulate the output voltage of the DAB to its desired value by controlling the phase shift via switching of two H-bridges \cite{dab,dab2}. The state-space representation of this stage can be written as
\begin{align}
\dot{v}_{hi}=&\frac{v_{fi}-v_{hi}}{C_{hi}r_{hi}}- \frac{n_{si}\phi_{si}(1-\phi_{si})}{2C_{hi}f_{i}L_{si}}v_{li}, \label{x4}\\
\dot{v}_{li}=&\frac{n_{si}\phi_{si}(1-\phi_{si})}{2C_{li}f_{si}L_{si}}v_{hi}-\frac{I_{dabi}}{C_{li}}, \label{x5}
\end{align}
\begin{align}
\dot{\xi}_{4i}=&v_{li}^\star-v_{li}.\label{xi4}
\end{align}
The control input for DAB is the phase shift ratio $\phi_{si}$, designed using a PI controller as
\begin{equation}
\phi_{si}=k_{7i}(v_{li}^\star-v_{li})+k_{8i}\xi_{4i}. \label{d3}
\end{equation}
\subsection{DC Distributed Renewable Energy Sources } 
The wind and solar generators, connected to the DC bus, are modeled as  current sources $I_w(t)$ and $I_{pv}(t)$ as shown in Fig 2. Similarly, the load is modeled as constant current $I_l(t)$. 
Further details on  dynamic modelling of  renewable sources in a SST-based microgrid appear in \cite{khan}.
\subsection{DC Distributed Energy Storage Devices (DESD)}
\begin{figure}[t!]
	\centering
	\includegraphics[width=0.84\columnwidth]{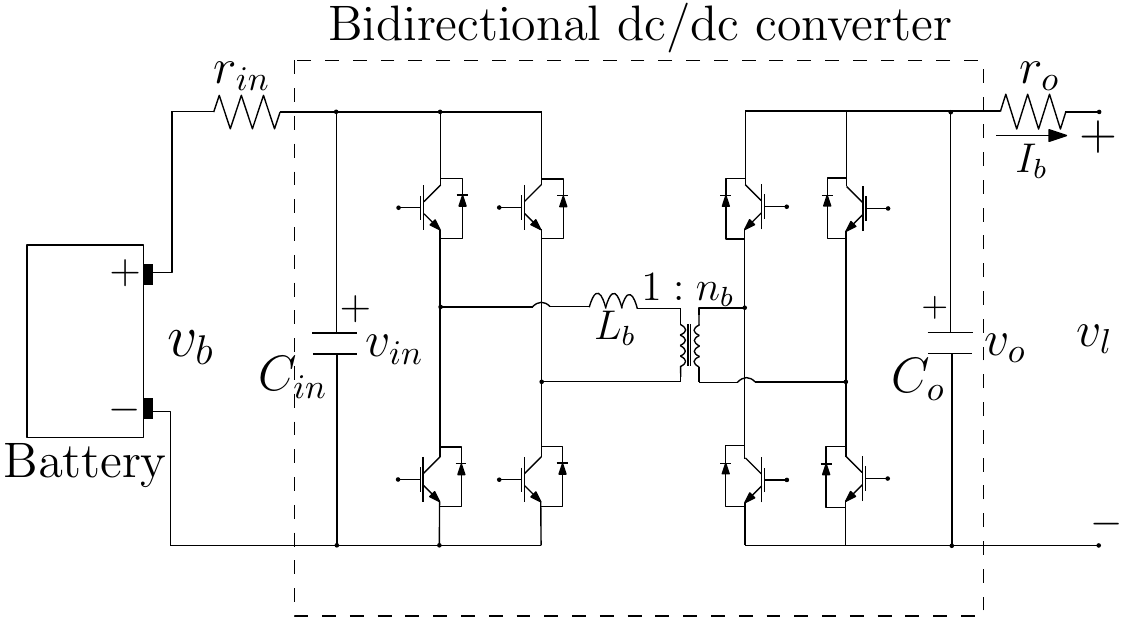}
	\caption{DC DESD circuit.\label{dcdesd}}
\end{figure}

A DESD is made of two interconnected units: a battery energy storage and an interfacing power converter. The battery is a complex electrochemical system whose equivalent electrical circuit models are widely available in the literature (see for example \cite{desd} and the references therein). Since, the battery can consume or dispatch power, a bidirectional DC-DC converter is needed to integrate the storage to the DC bus. The DAB converter in \eqref{x4}-\eqref{x5} can be used for this task. The resulting DESD system is depicted in Fig. \ref{dcdesd}. Its model equations are
\begin{align}
\dot{v}_{o}&=\frac{1}{r_{o}C_{o}}\left(v_{l}-v_{o}\right)-\frac{u_{b}}{C_{o}}v_{in},\label{z2b}\\
\dot{v}_{in }&=\frac{v_{b}-v_{in}}{C_{in}r_{in}}-\frac{u_{b}}{C_{in}}v_{o}\label{desd1}
\end{align}
where
\begin{equation}\label{ud}
u_{b}:=\frac{\phi_{b}(1-\phi_{b})n_{b}}{2f_bL_b}.
\end{equation}
and $\phi_b$ is the phase shift ratio which acts as the control input of the converter. The input signal $\phi_b\in[-1,1]$, i.e., it has a limited range of operation. Equations (15)-(17) denote the DESD model for the $i^{th}$ SST, but for simplicity we have dropped the subscript $i$. This slight abuse of notation will be followed in the forthcoming sections also whenever the subscript is clear from the context to avoid overuse of notations.

The output current of the storage $I_{b}$, shown in Fig. \ref{fig1} and \ref{dcdesd}, can be written as 
\begin{align}
I_{b}=&\frac{1}{r_o}(v_{o}-v_{l}). \label{ibat}
\end{align}
\noindent The battery voltage $v_b$ is always positive. The battery management system (BMS) supervises the appropriate operation of the battery currents and voltages. Protection hardware limits the battery operation to avoid damage of the equipment when is required by the BMS.  Thus, for the rest of the paper we make the following practical assumption:
\begin{assumption} \label{fa1} Voltage $v_b\in\mathbb{R}_{>0}\cap\mathcal{L}_{\infty}$ with  $v_b\in [v_b^{ min}, v_b^{ max}]$.\end{assumption}
\vspace{0.2cm}
\indent In the following sections, an isolated DC-DC converter will be used as the interfacing device between the voltage $v_l$ and the voltage of the DC-link for simplicity. An analogous control design and stability proof can be derived if a non-isolated converter is used.

\section{DESD Control Design and Stability Proof}\label{control}
\begin{figure}[t!]
	\includegraphics[width=\columnwidth]{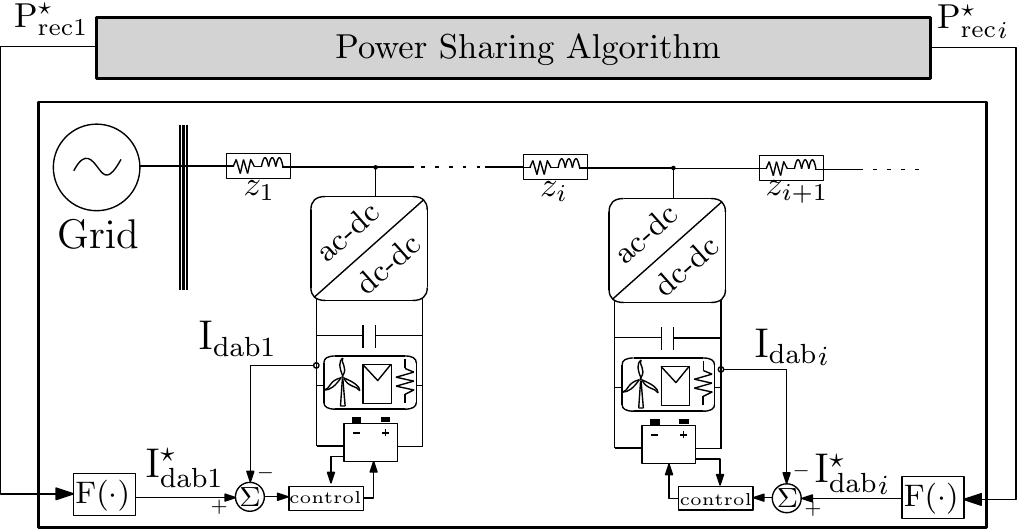}
	\caption{Multi-SST power sharing and energy storage control}\label{diag}
\end{figure}

A schematic digram for the operation of the overall networked microgrid is shown in Fig. \ref{diag}. 
In practice, a supervisory controller predicts the load for each microgrid $15$-$20$ minutes ahead of time, solves power flow, and generates the power setpoints $P_{rec}^\star$ for each SST.  However, problem arises when an unpredicted load change occurs  in between the scheduled instants. When that happens, the storage unit of the SST must trigger to produce or absorb the deficit power so that the power balance is maintained. The  setpoints in this case are calculated in accordance with the capacity of every storage unit from a power sharing algorithm  which will be provided shortly in Section \ref{powerd}.  

By regulating $I_{dab}$, the  proposed design permits to control the power flow between the microgrid and the transmission grid. This can be viewed as tertiary control following \cite{intro}, or secondary control following \cite{seco}. Once $P_{rec}^\star$ is scheduled by the power sharing algorithm, the control objective is to regulate the DC grid current $I_{dab}$ to
\begin{equation}\label{idabstar}
I_{dab}^\star =\frac{1}{v_l^\star}\left[P_{rec}^\star-\frac{(v_f^\star-v_h^\star)^2}{r_h}\right].
\end{equation}
\noindent This function is denoted as the `$\mathrm{F(\cdot)}$' block in Fig. \ref{diag}. Using KCL from Fig. \ref{fig1},
\begin{equation}\label{ld}
I_{dab}(t)=I_{pv}(t)+I_{w}(t)+I_{b}(t)-I_{l}(t)
\end{equation}
which in  steady-state operation becomes
\begin{equation}\label{ld2}
I_{dab}^\star=I_{pv}(t)+I_{w}(t)+I_{b}^r(t)-I_{l}(t),
\end{equation}
where $I_{b}^r(t)$ is the storage output reference current  needed to maintain $I_{dab}=I_{dab}^\star$. Subtracting  \eqref{ld2} to \eqref{ld} yields
\begin{equation}\label{ref_b}
I_{b}^r(t)=I_{b}(t)-I_{dab}(t)+I_{dab}^\star.
\end{equation}
\noindent Thus, the controller objective is:
\begin{itemize}
	\item[$\mathrm{C1}.$] To drive the current $I_b(t)$ to $I_{b}^r(t)$---see Section \ref{desd}.
	\item[$\mathrm{C2}.$] To guarantee stable operation of the entire SST network, i.e., the interconnected dynamic system \eqref{x1}-\eqref{ud} for $i=1,..,n$---see Section \ref{cont2}.
\end{itemize}
\vspace{-0.2cm}
\subsection{Control design}\label{cont1}
\noindent The controller design is presented in this section based on input-output (partial) linearization \cite{isidori}. 
As a first step, we define the tracking error $\delta$ 
\begin{align}
	\delta=&I_{b}-I_{b}^r
	=\frac{v_{o}-v_{l}}{r_{o}}-I_{b}^r \label{err}.
\end{align}
Then, using \eqref{desd1} the time-derivative of $\delta$ is
\begin{align}
	\dot \delta=&\frac{1}{r_o}(\dot v_{o}-\dot  v_{l}) -\dot I_{b}^r\nonumber\\
	=&\frac{1}{r_o^2C_o}(v_{l}-v_{o})+\frac{1}{r_oC_o}v_{in}u_{b}-\frac{\dot v_{l}}{r_o}-\dot I_{b}^r.\label{e_b}
\end{align}
  To assign an exponentially convergent error in closed-loop, the right-hand side of the last expression is equated to a scaling term of $\delta$: $$-\frac{\kappa_p}{r_oC_o}\delta=-\frac{\kappa_{p}}{r_oC_o}\left(\frac{v_o-v_l}{r_o}-I_{b}^r\right),$$  where (free) gain $\kappa_{p}>0$.  Doing so, one gets
\begin{align*}
	\frac{1}{r_{o}^2C_{o}}(v_{l}-v_{o})&+\frac{1}{r_{o}C_{o}}v_{in}u_{b}\\&\hspace{0.7cm}-\frac{\dot v_{l}}{r_o}-\dot I_{b}^r=-\frac{\kappa_{p}}{r_{o}C_{o}}\left(\frac{v_o-v_{l}}{r_o}-I_{b}^r\right).
\end{align*}
From the last equation
\begin{align} 
	u_{b}=&\frac{1}{v_{in}}\left[\frac{1}{r_o}(1-\kappa_{p})(v_o-v_l)+\kappa_{p}I_{b}^r\right.\nonumber\\&\hspace{3.5cm}\left.+r_oC_o\dot I_{b}^r+C_o\varphi_{v_{l}}(t)\right],\label{u_d}
\end{align}
with function $\varphi_{v_{l}}(t):=\dot v_{l}(t)$, i.e., where the expression for $\dot{v}_l$ follows from the RHS of \eqref{x5}.\\
\indent Note that although $u_b$ is chosen as the designable control input in \eqref{u_d}, the actual control input to the system is the phase shift $\phi_b$. Thus, with $u_d$ as in \eqref{u_d}, it is necessary to obtain the inverse mapping of \eqref{ud}. Since the duty cycle $u_b$ is always bounded within $[-1, 1]$, from the quadratic equation \eqref{ud} it follows that $\phi_b$ is given by the piecewise function 
\begin{equation}\label{d_i}
\phi_b=\left\{
\begin{array}{ll}
-\frac{1}{2}\pm\frac{1}{2}\sqrt{1-4h}&u_{b}\in[0,\frac{n_b}{8f_bL_b}]\\
-\frac{1}{2}-\frac{1}{2}\sqrt{1-4h}&u_{b}\in[-\frac{n_b}{f_bL_b},0],
\end{array}
\right.
\end{equation}
with
$h:=\frac{2f_{b}L_b}{n_b}u_b. $
\subsection{Stability of the DESD system}\label{desd}
\begin{figure}[t!]
	\centering
	\includegraphics[width=0.88\columnwidth]{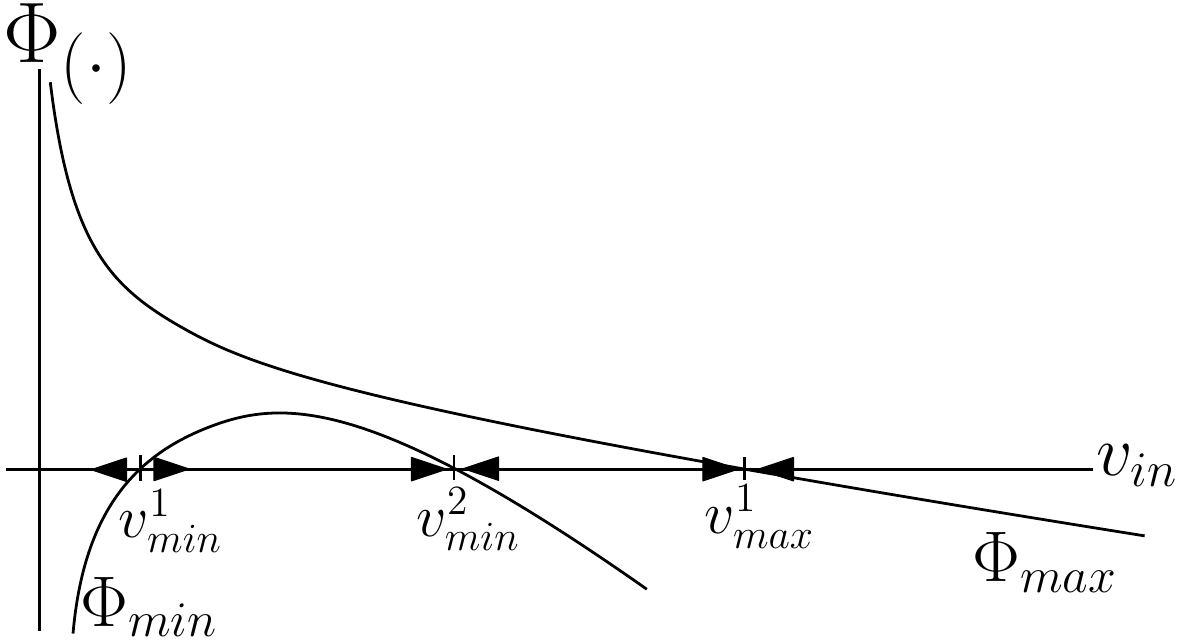}
	\caption{Plot of $v_{in}$ vs. $\Phi_{ max},\Phi_{ min}$}\label{plotvin}
\end{figure}

\begin{figure}[t!]
	\centering
	\includegraphics[width=\columnwidth]{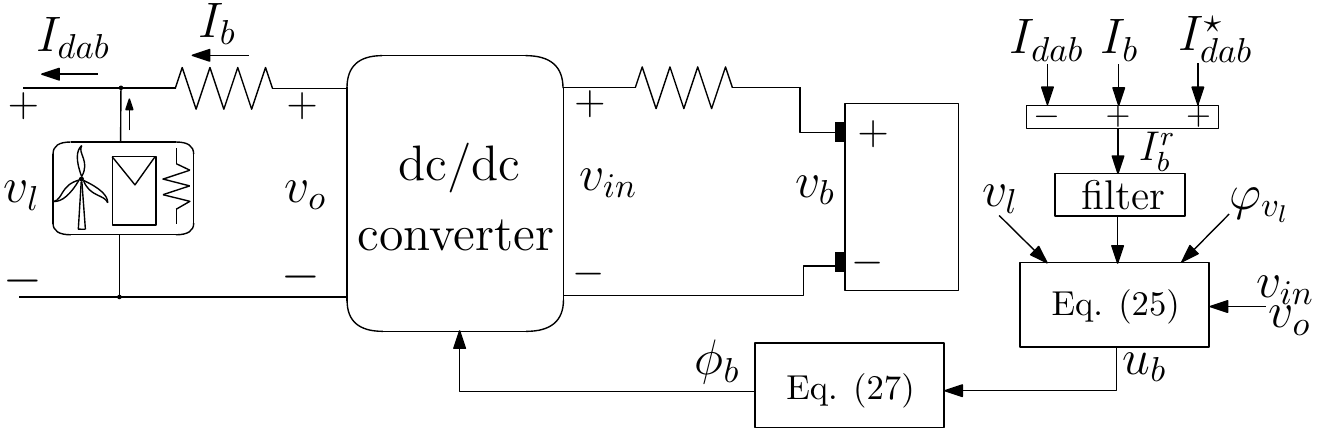}
	\caption{Implementation diagram for the proposed DESD controller}
	\label{imp}
\end{figure}

Note that the DESD system \eqref{z2b}-\eqref{desd1} has relative degree one with respect to $\delta$, i.e., the system is not fully feedback linearizable. Therefore,  stability of the residual dynamics has to be proven. This is addressed in the next proposition. Before proceeding,we make the following practical assumption.
\begin{assumption}\label{fact2}
	Signals $I_{b}^r$ and $\dot I_b^r$ are bounded and, for any operation around the equilibrium point, $v_l$ and $\varphi_{v_{l}}$ are also bounded for all $t\geq 0$.
\end{assumption}
Boundedness of $v_l$ and $\varphi_{v_{l}}$ will follow from the stability proof of the multi-SST system, provided in Section \ref{cont2}. At this point this is taken as an assumption to proceed with Proposition \ref{p}. Boundedness of the generation and load currents $I_w,I_{pv}$ and $I_l$ follows from the  physics. Thus, using \eqref{ld2}, $I_{b}^r$ is also bounded. Finally, since high frequencies in $I_b^r$ are filtered out when implementing (see Fig. \ref{imp}),  it is reasonable to consider $\dot I_{b}^r$ to be bounded as well.
\begin{proposition}\label{p}
	Consider the system \eqref{z2b}-\eqref{desd1} in closed-loop with \eqref{u_d}. Define $$	p(t):=v_o\left[\frac{1}{r_o}(1-\kappa_{p})(v_o-v_l)+\kappa_{p}I_{b}^r+r_oC_o\dot I_{b}^r+C_o\varphi_{v_{l}}\right].$$
	 Then,
	\begin{itemize}
		\item[P1.]The tracking error $\delta$ is exponentially stable with $$\delta(t)=\lim_{t\to\infty}\delta(0)\exp\left(-\frac{\kappa_p}{r_oC_o}t\right).$$ 
		\item[P2.]There exists a bound $|p(t)|\leq p_{ max}$ such that for all initial conditions $(v_o(0),v_{in}(0))\in\mathbb{R}^{2}$ with  $$v_{in}(0)>\frac{1}{2}v_b^{ min}-\frac{1}{2}\sqrt{{(v_b^{ min})}^2-4C_{in}p_{ max}}$$ 
		with ${(v_b^{ min})}^2-4C_{in}p_{ max}\geq 0$, 
		voltages $v_o$ and $v_{in}$ remains bounded. Moreover, there exist parameters $f_b,L_b$ and $n_b$ such that
		\begin{equation}\label{condition}
		u_b\in\left[-\frac{n_b}{f_bL_b},\frac{n_b}{8f_bL_b}\right].
		\end{equation}
	\end{itemize}
\end{proposition}
\vspace{0.3cm}
\textit{Proof.}
The closed-loop system is formed by the error dynamics together with \eqref{desd1} taking $u_b$ as in \eqref{u_d}. Thus, 
\begin{align}
\dot \delta=&-\frac{\kappa_{p}}{r_oC_o}\delta\label{error1}\\
\dot  v_{in} =&-\frac{v_{in}}{r_{in}C_{in}}- \frac{p(t)}{C_{in}v_{in}}+\frac{v_b}{r_{in}C_{in}}.\label{error2}
\end{align}
Since \eqref{error1} is decoupled from \eqref{error2}, P1 immediately follows. Next, using  \eqref{err} one can conclude boundedness of $v_{o}$ and,  from Assumption \ref{fact2}, the existence of bound $p_{ max}$ in $p(t)$. 
For $v_{in}>0$,
\begin{align*}
\Phi_{ min}(v_{in})\leq \dot v_{in}\leq \Phi_{ max}(v_{in})
\end{align*}
where 
\begin{align}
\Phi_{ min}(v_{in}):=&-\frac{v_{in}}{r_{in}C_{in}}-\frac{p_{ max}}{C_{in} v_{in}}+\frac{v_{b}^{ min}}{r_{in}C_{in}}\label{vmin}
\end{align}
\begin{align}
\Phi_{ max}(v_{in}):=&-\frac{v_{in}}{r_{in}C_{in}}+\frac{p_{ max}}{C_{in} v_{in}}+\frac{v_{b}^{ max}}{r_{in}C_{in}}\label{vmax}
\end{align}
and constants $v_b^{min},v_b^{max}> 0$ follows from Assumption \ref{fa1}.
\noindent The corresponding plots are depicted in Fig. \ref{plotvin}. The points  $v_{{ min}}^1$ and $v_{ min}^2$ correspond to the roots of \eqref{vmin}, and $v_{{ max}}^1$ corresponds to the positive root of \eqref{vmax}. Thus,
\begin{equation}
\begin{aligned}\label{roots}
v_{ min}^{1,2}=&\frac{1}{2}v_b^{ min}\pm\frac{1}{2}\sqrt{{(v_b^{ min})}^2-4C_{in}p_{ max}}\\v_{ max}^1=&\frac{1}{2}v_b^{ min}+	\frac{1}{2}\sqrt{{(v_b^{ min})}^2+4C_{in}p_{ max}}.
\end{aligned}
\end{equation}
\noindent Now, consider the following two models 
\begin{equation}
\begin{aligned}
\label{lbound}\dot v_{ min}=\Phi_{ min}(v_{ min}),\;\; \dot v_{ max}=\Phi_{ max}(v_{ max}).
\end{aligned}
\end{equation}
\noindent From the Comparison Lemma \cite{khal}, for all $v_{in}(t)>0$,   $$v_{ min}(t)\leq v_{in}(t)\leq v_{ max}(t).$$ 
For all $v_{min}(0)> v_{ min}^1$, $v_{ min}(t)$ remains positive and converges to $v_{ min}^2$. In the same way,  for all $v_{max}(0)>0$, $v_{ max}(t)$ stays positive and converges to $v_{ max}^1$. This proves P2. This also shows that $v_{in}$ only takes strictly positive values, i.e., it is never zero, which in turn proves that $u_b$  is bounded. Therefore, there exist free parameters satisfying \eqref{condition}. \hfill{$\blacksquare$}\vspace{0.3cm}\\
The physical interpretation of the boundedness of $v_o$ and $v_{in}$ as proved above can be explained as follows. With $I_b$ bounded, the output power $p_{o}=v_lI_b-r_{o}I_{b}^2=v_oI_b$ is bounded and, thus, $v_o$ is also bounded. Since the  converter is a passive device, it follows that the input power $p_{in}=\frac{v_b-v_{in}}{r_{in}}v_{in}$ is bounded and, thus, $v_{in}$ is also bounded. 

 The controller implementation diagram is shown in Fig. \ref{imp}. Notice that, to calculate $I_{b}^r$, measurements of currents $I_{dab}$ and $I_{b}$ are required. As mentioned, the signal $I_{b}^r$ is passed through a low-pass filter to eliminate noise. The phase shift then follows from  \eqref{d_i}. Also, bounds \eqref{condition} in Proposition  \ref{p} imply that by choosing appropriately the system parameters $n_{b}, L_{b}$ and $f_{b}$, the control input $\phi_b$ can be ensured to be unsaturated.  The parameters can be chosen by the designer to enhance the transient performance of the battery states. 

\subsection{Stability of the $n$-SST System}\label{cont2}
\begin{figure}[t!]
	\centering
	\includegraphics[width=1\columnwidth]{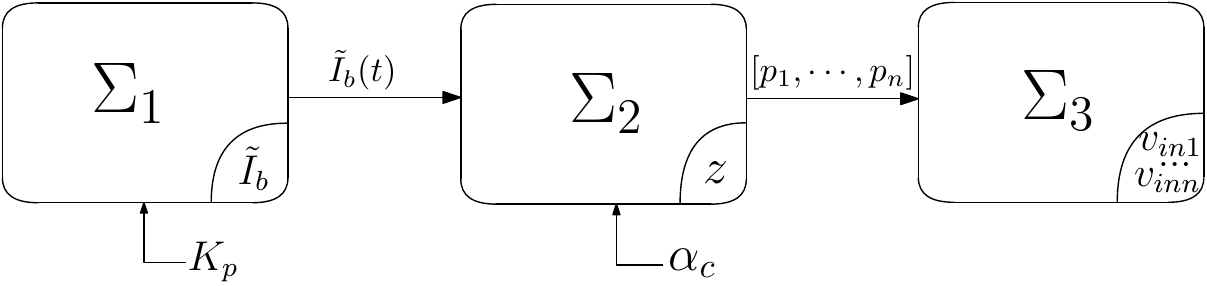}
	\caption{Cascade representation of the closed-loop system.}
	\label{repr}
\end{figure}

We next extend the stability proof to a $n$-SST network where $n\geq1$. Following standard assumptions as in \cite{eric}, we neglect the effect  of the second harmonics on the dynamics of $v_{fi}$ in \eqref{vf} as their impact on the steady-state value of (3) is small from small-ripple approximation. Since we are primarily interested in the fundamental frequency response, the $v_{fi}$-dynamics is approximated to
\begin{align}
\dot v_{fi}=&-\frac{v_{fi}-v_{hi}}{C_{fi}}-\frac{1}{2C_{fi}}(d_{1i}i_{di}+d_{2i}i_{qi})\label{mx3}.
\end{align}
\noindent Using \eqref{ref_b} and the definition of $\delta$, we rewrite \eqref{x5} in the equivalent form 
\begin{equation}
\dot{v}_{li}=\frac{n_{si}\phi_{si}(1-\phi_{si})}{2C_{li}f_{si}L_{si}}v_{li}-\frac{I_{dabi}^\star}{C_{li}}-\frac{\delta_{i}}{C_{li}}.\label{mx5}
\end{equation}
The closed-loop is then conformed by equations  \eqref{x1},  \eqref{x2}, \eqref{mx3}, \eqref{xi1}-\eqref{xi3}, \eqref{x4}, \eqref{mx5} and \eqref{error1}-\eqref{error2}. A block diagram representation is showed in Fig. \ref{repr}. As it can be seen, the closed-loop admits a (double) cascade representation. The cascaded system $\Sigma_1-\Sigma_2$ is given by
\begin{align}\label{twosst}
	\begin{bmatrix}
		\dot z\\\dot {\tilde I}_b
	\end{bmatrix}=\begin{bmatrix}\gamma(z,\alpha_c)\\K_p\tilde I_b\end{bmatrix}+\begin{bmatrix}
		P\\0\end{bmatrix}\tilde I_b
\end{align}
where the function $\gamma:\mathbb{R}^{9n}\to\mathbb{R}^{9n}$ accounts for the nonlinearities of the rectifier, DAB converter and controllers dynamics. Also,  $z^\top=[z_1^\top~\cdots~z_n^\top]\in\mathbb{R}^{9n}$. The constant vector $\alpha_c\in\mathbb{R}^{8n}$ contains the controller gains. The subvectors $z_i^\top=[{x_i}^\top\;~{\xi_{1i}}~\;\cdots\; {\xi_{4i}}]$ with ${x_i^{\top}}=[i_{di}~i_{qi}~v_{fi}~v_{hi}~v_{li}]$. We define vector $\tilde I^\top_b=[\delta_1~\cdots~\delta_n]\in\mathbb{R}^n$,  and constant matrices $P=[-\frac{1}{C_{f1}}v_5\;\cdots\;-\frac{1}{C_{fn}}v_{(9n-4)}]\in\mathbb{R}^{9n\times n}$  and  $K_p=\mathrm{diag}(-\frac{\kappa_{p1}}{r_{o1}C_{on}},\cdots,-\frac{\kappa_{pn}}{r_{on}C_{on}})\in\mathbb{R}^{n\times n}$. We have denoted  by $v_j\in\mathbb{R}^{9n}$ a vector of the Euclidean basis with its  $j$th-element equal to one. The linearized state model of \eqref{twosst} can be the written as
\begin{equation}\label{A}\begin{bmatrix}\dot {\delta z}\\\dot {\tilde{I}}_b\end{bmatrix} = \begin{bmatrix}\Gamma(\alpha_c)&P\\0&K_p\end{bmatrix}\begin{bmatrix}\delta z \\\tilde I_b\end{bmatrix}\end{equation}
where $\Gamma\in\mathbb{R}^{9n\times 9n}$ is 
$$\Gamma=\frac{\partial\gamma(z)}{\partial z}\rvert_{z=z_\star},$$ with $z_\star$ the equilibrium point of \eqref{x1}-\eqref{ud} for $i=1,\cdots ,n$. Note that the eigenvalues of  \eqref{A} are  $\mathrm{eig}\{\Gamma\}\cup\{-\frac{\kappa_{p1}}{r_{o1}C_{o1}},\cdots,-\frac{\kappa_{pn}}{r_{on},C_{on}}\}$. Thus, if every PI gain (i.e., vector $\alpha_c$) is selected such that $\mathrm{Re}[\mathrm{eig}\{\Gamma\}]<0$, then the closed-loop system of interconnected $n$ SST-driven microgrids \eqref{x1}-\eqref{ud} will be locally asymptotically stable.

Block $\Sigma_3$ in  Fig. \ref{repr}  corresponds to the $v_{ini}$-dynamics ($i=1,\cdots,n$) introduced in \eqref{error2}. From stability of the cascade $\Sigma_1-\Sigma_2$, it follows that functions $p_i$, the inputs of Block $\Sigma_3$, are bounded. The later validates Assumption \ref{fact2} and  Proposition \ref{p}. Thus, the overall closed-loop system is stable.
\section{Multi-SST Power Sharing}\label{powerd}
Change in load currents can be categorized into two scenarios. The first scenario is when the change is small enough that $|I_{b}^r|\leq I_{b}^{max}$, i.e., the battery current magnitude of $I_{b}^r$ needed for the slack is less than the maximum capacity of the battery. In this case, the SST whose load changed can compensate for the deficit power locally using the decentralized controller \eqref{ud}, as described in Section III. However, if the load change is so high that $|I_{b}^r|\geq I_{b}^{max}$, then other SSTs in the network need to be used to support the deficit. The control action will still remain decentralized as in section \ref{control}, but the computation of the voltage and current setpoints for the $m^{th}$ SST (i.e., the SST whose load changed) will now depend on the neighboring SSTs. The way to implement this can be as follows. First, the inability of the $m^{th}$ SST in supporting its new load is detected, and immediately after the setpoints $P^\star_{reci}$ of every SST, $i=1,..,n$, are overwritten to enable power sharing. Subsequently, the corresponding currents setpoints $I_{dabi}^\star$ are updated according to \eqref{idabstar}. For any iteration,  $P_{reci}^\star$ must satisfy the power balance equation---for details on this equation, please see \cite{ali}
\begin{equation}
	\left(i_{di}^\star+\frac{v_{di}^\star}{2r_{fi}}\right)^2+\left(i_{qi}^\star+\frac{v_{qi}^\star}{2r_{fi}}\right)^2=\frac{v_{di}^{\star 2}+v_{qi}^{\star 2}}{4r_{fi}^2}-\frac{2P_{reci}^\star}{r_{fi}}.\label{multicirc1}
\end{equation}

\indent Any change in $i_{di}^\star$ or $v_{di}^\star$ in one of the SST in the network will also impact the others. In order to avoid this complexity power sharing methods for SST systems developed in our previous work \cite{ali} are utilized in a cooperative multi-SST context. These algorithms allow to properly update $i_{di}^\star,i_{qi}^\star, v_{di}^\star$ and $v_{qi}^\star.$ They are summarized below.
\subsection{Method 1:  Constant Current Method}
\begin{algorithm}[t!]
	\caption{Constant current method}\label{a1}
	\small{
	\begin{algorithmic} 
		\WHILE {$I_{bm}^r=I_{bm}^{max}$}
		\IF{$\delta_{m} > 0$}
		\STATE $P_{recm} \leftarrow$ Calculate the updated power requirement based on equation (19).
		\IF{$P_{recm}\leq P_{recm}^{max}$}
		\STATE $P_{recm}^\star \leftarrow P_{reci}$
		\ELSE
		\STATE $P_{recm}^\star \leftarrow P_{recm}^{max}$
		\ENDIF
		\STATE $i_{dm}, v_{dm} \leftarrow$ Update following equation (38) for the $m^{th}$ SST
		\STATE $v_{di} \leftarrow$ Update following Power Sharing Method 1; $i=1\to n, i\neq m$
		\STATE $P_{reci}\leftarrow$ Update using equation (39) for the $i^{th}$ SSTs
		\IF{$P_{reci}\leq P_{reci}^{max}$}
		\STATE $P_{reci}^\star \leftarrow P_{reci}$
		\ELSE
		\STATE $P_{reci}^\star\leftarrow$ Continue the IEM command
		\STATE $P_{recm}^\star\leftarrow$ Continue the IEM command
		\ENDIF
		\ENDIF
		\ENDWHILE
	\end{algorithmic}}
\end{algorithm}
 In this method,  when there is a load change in the $m^{th}$ SST that cannot be supported by its local storage, the other SSTs assist without changing its input current. That means that for all $i\not=m$, the currents $i_{di}^\star$ and $i_{qi}^\star$ remain unaltered. The input voltage of all the SSTs is updated to a new value as a function of the change in power. The analytical expressions for the change of the input current of the $m^{th}$ SST can be expressed as
\begin{align}
	\left(r_{fm}+\sum_{k=1}^{m}r_{k}\right)&\Delta{{i}_{dm}^{\star 2}}+\left(2r_{fm}i_{dm}^\star+v_{dm}^\star+\sum_{k=1}^{i}r_{k}i_{dm}^\star\right.\nonumber\\&\hspace{-0.8cm}\left.+\sum_{k=1}^{i}x_{k}i_{qm}^\star\right)\Delta{i}_{dm}+2\Delta{P_{recm}}=0.\label{multi6}
\end{align}
\noindent Keeping $i_{qm}^\star$ unchanged, the current change $\Delta i_{dm}^\star$ of the $m^{th}$ SST  is the root of \eqref{multi6}. Also, the voltage relation follows as
$$\Delta{v_{dm}}+j\Delta{v_{qm}}=(\Delta{i_{dm}}+j\Delta{i_{qm}})\sum\limits_{k=1}^m (r_k+jx_k).$$
 The voltage drop at all other SSTs in the network will be the same as that of the $m$$^{th}$ SST to maintain the input current as it is. Algorithm 1 shows step by step execution of Method 1 to update the new power setpoints in a radial network. 
 Once the power setpoints $P_{reci}^\star$ are updated, $I_{dabi}^\star$ is recalculated as in equation (\ref{idabstar}). The battery current is then regulated with the designed local controller in Section \ref{control} with updated $I_{dabi}^\star$.
\subsection{Method 2: Constant Voltage Method}
The second method maintains feasible operation through constant node voltage of all other SSTs when there is any change in the $m^{th}$ SST. Because of the radial configuration, only the input current references of the immediate neighboring SSTs: $(m-1)^{th}$ and $(m+1)^{th}$ SST change whereas the other setpoints remain invariant \cite{ali}. Algorithm 2 shows step by step execution of method 2 to update the power setpoints of the neighboring SSTs in a radial network. Once the power setpoints $P_{reci}^\star$ are updated for the neighbors, $I_{dabi}^\star$ are recalculated as in equation (\ref{idabstar}) and battery current is subsequently regulated with the designed local controller in Section \ref{control} following updated $I_{dabi}^\star$. Note that in this method, only the immediate neighbors share the load which may not be practical for a big change in the load power. This method is thus more suitable for smaller-scale microgrid networks where every SST may not have storage.
\begin{algorithm}
	\caption{Constant voltage method}\label{a2}
	{\small
		\begin{algorithmic} 
			\WHILE {$I_{bm}^r=I_{bm}^{max}$}
			\IF{$\delta_{m} > 0$}
			\STATE $P_{recm} \leftarrow$ Calculate the updated power requirement based on equation (19).
			\IF{$P_{recm}\leq P_{recm}^{max}$}
			\STATE $P_{recm}^* \leftarrow P_{recm}$
			\ELSE
			\STATE $P_{recm}^* \leftarrow P_{recm}^{max}$
			\ENDIF
			\STATE $i_{dm}, v_{dm} \leftarrow$ Update following equation (39) for $m^{th}$ SST
			\STATE $i_{di} \leftarrow$ Update following Power Sharing Method 2; $i=(m-1),(m+1)$
			\STATE $P_{reci}\leftarrow$ Update using equation (38) for $i^{th}$ SSTs
			\IF{$P_{reci}\leq P_{reci}^{max}$}
			\STATE $P_{reci}^* \leftarrow P_{reci}$
			\ELSE
			\STATE $P_{reci}^*\leftarrow$ Continue the IEM command
			\STATE $P_{recm}^*\leftarrow$ Continue the IEM command
			\ENDIF
			\ENDIF
			\ENDWHILE
	\end{algorithmic}}
\end{algorithm}
\section{Simulation Results}\label{sim}
The proposed controller and power sharing methods are next validated using simulations on a radial 9-bus distribution feeder model containing one SST at each bus. The tie-line impedances of this model, which are based on the IEEE 34-bus distribution system, are:  $Z_{01}= 0.653+j0.651,~Z_{12}=0.438+j0.437,~Z_{23}=8.16+j8.14,~Z_{34}=9.49+j9.47,~Z_{45}=7.53+j7.51,~Z_{56}=0.0037+j0.0027,~Z_{56}=0.0037+j0.0027,~Z_{67}=0.906+j0.481,~Z_{78}=25.52+j13.546,~Z_{89}=7.284+j13.865$. 
The SST models are identical and their parameters are based on the GEN-II SST model \cite{wang}. First, results are provided to validate the control with added stochastic randomness to the loads and current sources representing the wind and solar energy. Then, the designed controller is utilized to apply Method 1 in case of a sudden change in the load when the local storage is not capable to support the change. The latter is extended to the case when there exist delays in the computation of the setpoint updates. 
To avoid redundancy,closed-loop responses of only $\mathrm{SST1}$ through $\mathrm{SST3}$ are displayed in each figure.

\begin{figure}[h!]
	\centering
	\includegraphics[width=0.93\columnwidth]{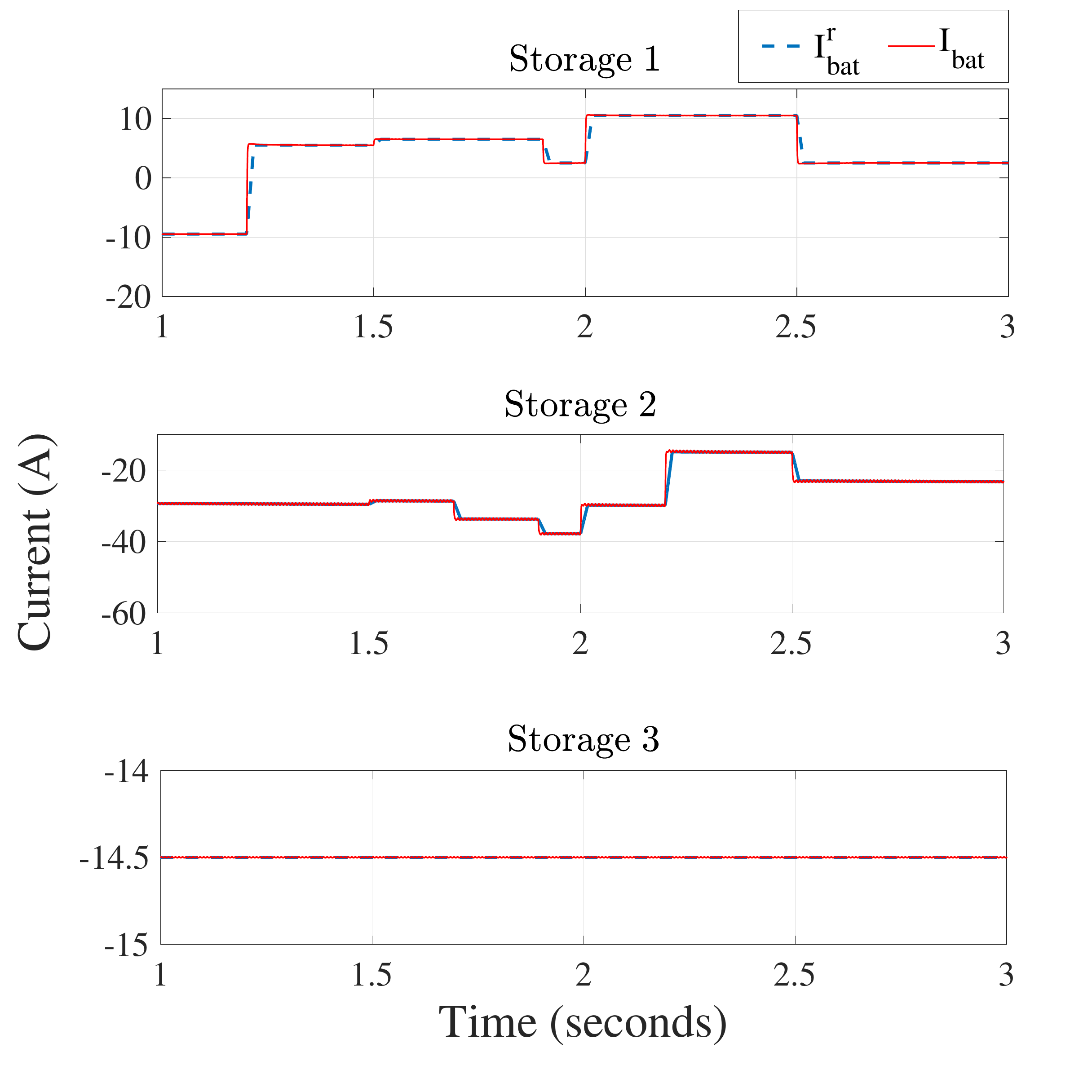}
	\caption{Storage current profile with change in net loads.}\label{sst_local}
\end{figure}
\begin{figure}[h!]
	\centering
	\includegraphics[width=0.93\columnwidth]{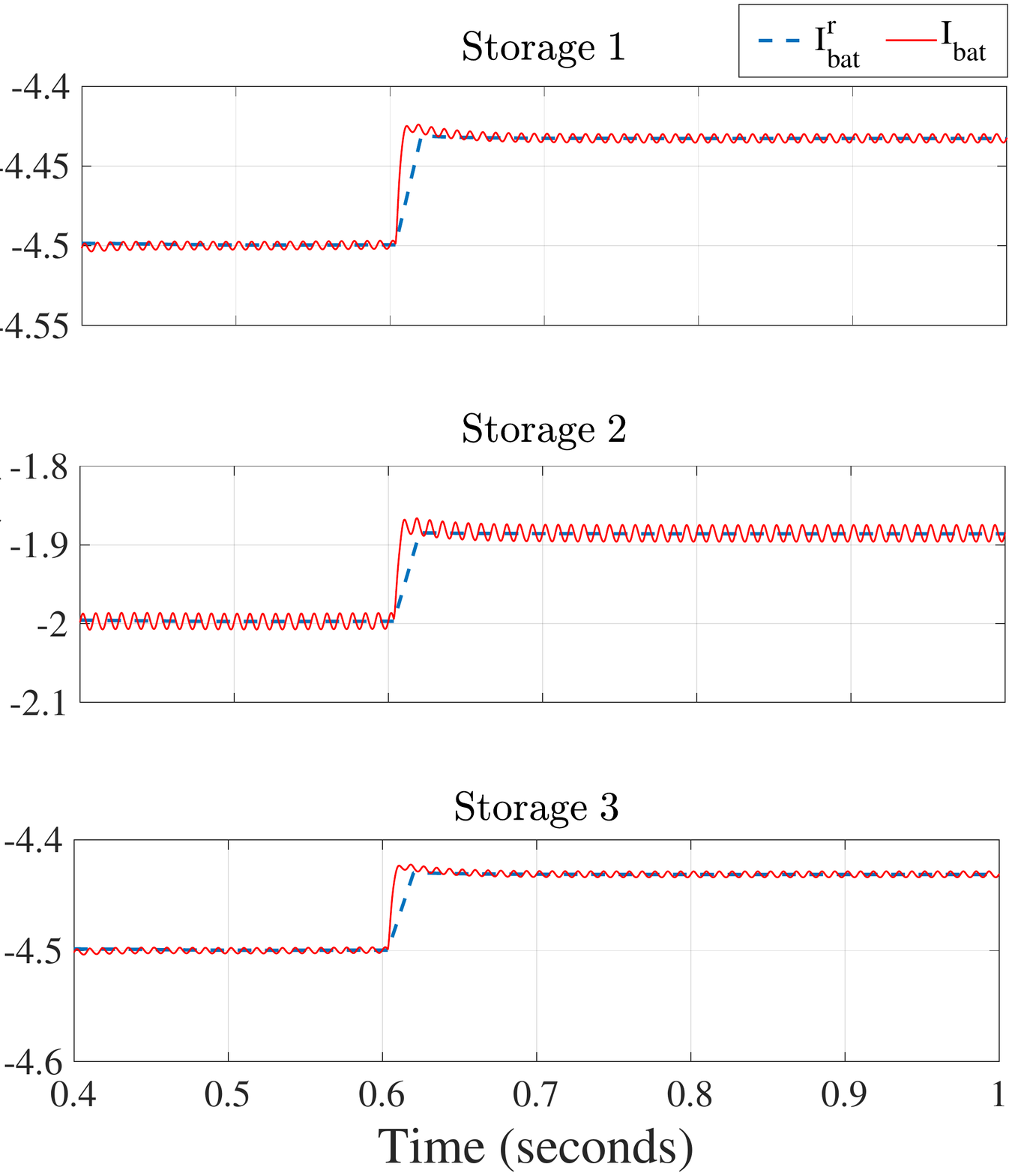}
	\caption{Storage current  profiles when  sharing power after a change in net load in $\mathrm{SST1}$.}\label{sst_ps10}
\end{figure}
\begin{figure}[h!]
	\centering
	\includegraphics[width=0.93\columnwidth]{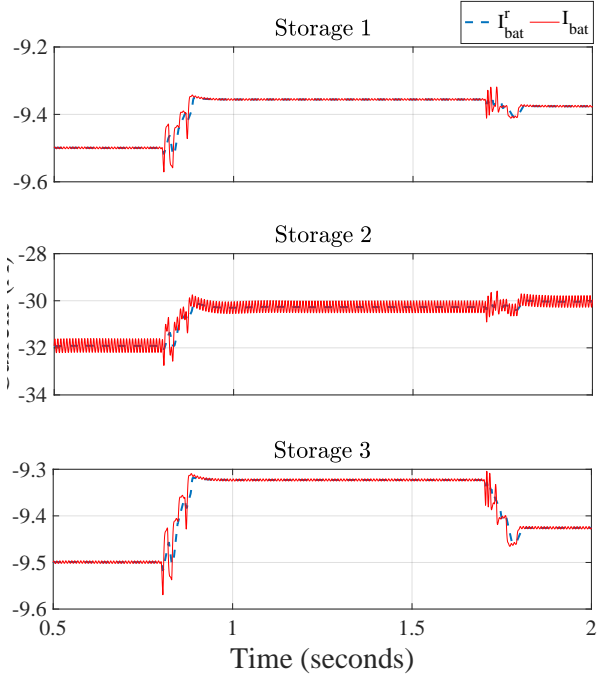}
	\caption{Storage current profile when sharing power delay after a change in net load in $\mathrm{SST1}$.}\label{sst_ps11}
\end{figure}
\subsection{Autonomous operation: $|I_{b}^r|\leq I_{b}^{max}$}
The simulation starts with a nominal load of 1 kW at time $t=0$ in $\mathrm{SST1}$ and $\mathrm{SST3}$ to $\mathrm{SST7}$. $\mathrm{SST2}$ starts with 10 kW and $\mathrm{SST8}$ and $\mathrm{SST9}$ with $-1$ $\mathrm{kW}$. Load and renewable generation are changed in all the SSTs except for $\mathrm{SST3}$, $\mathrm{SST5}$, $\mathrm{SST7}$ and $\mathrm{SST9}$. The storage response to the dynamic $I_{b}^r$ is observed in Fig. \ref{sst_local} for $\mathrm{SST1}$-$\mathrm{SST3}$. Similar responses were observed for $\mathrm{SST4}$-$\mathrm{SST9}$  as well. It is found for all the SSTs, the storage current accurately follows the reference (dotted line) for any changes in the load (as long as ${I_{b}^r}\leq{I_{b}^{max}}$) with the developed controller. $I_{b}^{max}=50\;\mathrm{A}$ for simulation validation. The net load is varied from $0\;\mathrm{A}$ to $10\;\mathrm{A}$ by controlling PV current level  and load demands. When there is no change in the net load then $I_{b}^r$ remains as unchanged. 
\begin{table}[t]
	\centering
	\caption{Power setpoints  before/after the load change ($\mathrm{kW}$).}\label{table}
	\begin{tabular}{cccc}\toprule
		$\mathrm{SST\#}$&$P_{rec}^{\star}(t< t_0)$  &\multicolumn{2}{c}{$P_{rec}^{\star}(t\geq t_0)$ }\\
		&&$\mathrm{Method\;1}$&$\mathrm{Method\;2}$\\
		\midrule
		$1,3,8,9$&$-1$&$-1.02$7&$-1$\\
		$2$&$-2$&$-2.044$&$-2$\\
		$4$&$1$&$1.06$&$0.95$\\
		$5$&$1$&$-0.6$&$-0.6$\\
		$6$&$1$&$1.06$&$2.53$\\
		$7$&$1$&$1.06$&$1$\\	\bottomrule
	\end{tabular}
\end{table}
\subsection{Power Sharing operation: $|I_{b}^r|>I_{b}^{max}$}
The wind and PV currents are kept constant for all the SSTs before $t_0=0.6$ $\mathrm{s}$. Power setpoints $P_{reci}^\star$ are shown in Table \ref{table}. Afterwards, a sudden load change of $0.5\;\mathrm{kW}$ in the net power happens at $t_0=0.6$ s in $\mathrm{SST5}$ which drives the magnitude of $I_b^r$ to exceed $I_b^{max}$ set at $12\;\mathrm{A}$. 
Subsequently, $I_{b}^r$ gets updated for all the SSTs and  $I_{b}$ follows as shown in Fig. \ref{sst_ps10}. The ripples present in the response of $I_{b}$ is due to the second harmonics of the rectifier output voltage \eqref{vf} that also impact the output voltage $v_{l}$ because of the battery interface into the DC bus. However, the ripples are very small in magnitude and within a range of $0.001\%-0.005\%$. The setpoint calculation for power sharing is almost instantaneous, and hence the control is actuated immediately once the new setpoints are calculated.The convergence times for the closed-loop response of the SST currents is around  $0.05$ $\mathrm{s}$.  To compare Method 1 with 2, Table \ref{table} shows the new setpoints $P_{rec}^{\star}(t\geq t_0)$ calculated with both algorithms. As seen from the table, 
in method 1, all the SSTs update their setpoint along with grid to support $\mathrm{SST5}$. On the other hand, in method 2, only its immediate neighbors compensate the load changes. It is important to mention that grid current remains unchanged in Method 2 as the change in power is fully compensated by its neighbors.  \\
\indent Finally, we simulate a scenario in which the operation points are not updated immediately after the load change. This is intended to emulate  the computational delay that exists while updating the new setpoints. The results are shown in Fig. \ref{sst_ps11}. Oscillations are observed as the SSTs operate under wrong setpoints. However, as the setpoins and  $I_{b}^r$ are updated within $0.01$ $\mathrm{s}$, the system reaches a new steady state within $0.05$ $\mathrm{s}$.
\section{Conclusion}\label{conclusi}
\noindent This paper developed a nonlinear control framework for controlling storage devices in networked microgrids considering the intermittent behavior of renewable energy sources and loads. Special attention is paid to the nonlinear dynamics of both the microgrid model and the storage model for designing this controller, and guaranteeing closed-loop stability. The controller can be implemented in a completely decentralized way using local output feedback only. Results are verified using IEEE prototype distribution feeder models. The proposed approach can be particularly important for power sharing among microgrids during storms and natural calamities, when power from healthy parts of the network need to be transfered to other remote parts in a stable way. Future work along this direction will include extension of these results under various cyber-physical uncertainties in SST-to-SST communication, and also evaluating the impact of malicious cyber intrusions (such as misleading manipulations in the SST setpoints) on closed-loop stability.  
\vspace{-0.21cm}
\bibliographystyle{IEEEtran}
\bibliography{bibliography}
\end{document}